\documentclass[11pt,letterpaper]{article}
\usepackage{amssymb}
\usepackage{amsfonts}
\usepackage{amsmath}
\usepackage[numbers,round]{natbib}
\usepackage{epsfig}
\usepackage{graphicx}
\usepackage[letterpaper]{geometry}

\def\keywords#1{
\par\vspace{0.5ex}{\noindent\normalsize\bf Keywords:} #1
\vspace{0.5ex}   
}
\def\@begintheorem#1#2{\par\bgroup{\sc #1\ #2. }\it\ignorespaces}
\def\@opargbegintheorem#1#2#3{\par\bgroup{\sc #1\ #2\ (#3). }\it\ignorespaces}
\def\@endtheorem{\egroup}

\begin{document}

\title{The evolution of virulence in RNA viruses under a
competition-colonization trade-off}
\author{Edgar Delgado-Eckert$^{1,2,}$\thanks{%
*Corresponding author. Email address: edgar.delgado-eckert@mytum.de} \and %
Samuel Ojosnegros$^{1}$ \and Niko Beerenwinkel\medskip $^{1,2}$ \and $^{1}$%
{\small {Department of Biosystems Science and Engineering, }ETH Zurich,{\
Mattenstrasse 26, 4058 Basel, Switzerland.}} \and $^{2}${\small Swiss
Institute of Bioinformatics.}}
\maketitle

\begin{abstract}
RNA viruses exist in large intra-host populations which display great
genotypic and phenotypic diversity. We analyze a model of viral competition
between two viruses infecting a constantly replenished cell pool, in which
we assume a trade-off between the virus' colonization skills (cell killing
ability or virulence) and its local competition skills (replication
performance within coinfected cells). We characterize the conditions that
allow for viral spread by means of the basic reproductive number and show
that a local coexistence equilibrium exists, which is asymptotically stable.
At this equilibrium, the less virulent competitor has a reproductive
advantage over the more virulent colonizer. The equilibria at which one
virus outcompetes the other one are unstable, i.e., a second virus is always
able to permanently invade. One generalization of the model is to consider
multiple viral strains, each one displaying a different virulence. However,
to account for the large phenotypic diversity in viral populations, we
consider a continuous spectrum of virulences and present a continuum limit
of this multiple viral strains model that describes the time evolution of an
initial continuous distribution of virulence. We provide a proof of the
existence of solutions of the model's equations and present numerical
approximations of solutions for different initial distributions. Our
simulations suggest that initial continuous distributions of virulence
evolve towards a stationary distribution that is extremely skewed in favor
of competitors. Consequently, collective virulence attenuation takes place.
This finding may contribute to understanding the phenomenon of virulence
attenuation, which has been reported in previous experimental studies.
\end{abstract}



\section{Introduction}

RNA viruses are fast evolving pathogens that can adapt to continuously
changing environments. Due to their error-prone replication, large
population size, and high turnover, RNA virus populations exist as
quasispecies (\citet{Holland1992,Domingo1997}). The viral mutant spectrum
consists of many genetic variants which give rise to diverse phenotypes.
This phenotypic diversity is reflected in different traits, including the
ability to kill host cells (herein referred to as virulence).

The concept of virulence has been used in various areas of the life sciences
with different meanings. For instance, in evolutionary biology the virulence
of a pathogen is defined as the fitness costs to the host that are induced
by the pathogen. In epidemiology the term usually means the pathogen-induced
host mortality. In more clinical settings virulence often refers to the
severity of disease symptoms induced by a pathogen. In this article, we
consider intra-host virus dynamics and use the term virulence to denote the 
\textit{cell killing rate} of a virus infecting tissue (%
\citet{EstebanDomingo}). Thus, we apply the epidemiological meaning of the
concept of virulence to the intra-host viral microepidemics. This use of the
concept of virulence is related to the macroscopic or inter-host
significations just explained, because, in general, the cell killing rate of
a virus affects the course of infection, its inter-host repercussions as
well as the mortality of the host.

The evolution of virulence has been vastly studied in experimental and
theoretical approaches in a variety of host-pathogen interactions and under
diverse conditions or assumptions (see, among many others, %
\citet{BullVirulenceAReview}, \citet{CoinfectionAndEvolutionVir}, %
\citet{LenskiIII}, \citet{LenskiII}, \citet{LenskiIV}, \citet{LenskiV}, %
\citet{LenskiI}, \citet{EvoVirII}, \citet{EvoVirI}).

Different mechanisms have been proposed for the frequently observed
coexistence of several viral strains. Quasispecies theory suggests that
viral variants are maintained in a mutation-selection equilibrium (%
\citet{Eigen1988}). In ecology, coexistence often results from spatially
structured habitats (\citet{Tilman2007}). A trade-off between the ability of
each individual to colonize unoccupied territory and to compete with others
for the same habitat patch can result in coexistence of two strategies:
competition and colonization. Competitors have an advantage when competing
locally for resources, whereas colonizers are more successful in reaching
new resources.

Competition-colonization dynamics in an RNA virus infection have been
recently demonstrated \textit{in vitro} by \citet{Ojosnegros2010} in an
experimental and theoretical approach. In this system, highly virulent viral
strains play the role of colonizers, because they kill cells faster and thus
replicate faster which allows faster spread and colonization of new cells.
Local competition arises when two or more different viruses infect the same
cell and compete for intracellular resources. Competitors manage to produce
more offspring in a cell coinfected together with a colonizer and, at the
same time, extend the cell killing time characteristic of a colonizer. This
phenomenon is termed viral interference.

The competition-colonization coevolutionary dynamics of two different viral
strains in cell culture have been described in \citet{Ojosnegros2010} by a
modification of the basic model of virus dynamics (%
\citet{Nowak2000,Perelson1999}). The model predicts that the outcome of
viral competition for the cell monolayer depends on the initial density of
viruses per cell. Under low initial density, colonizers produce more total
offspring, whereas under high-density conditions, coinfection is more likely
to occur and hence competitors have a selective advantage. This prediction
was confirmed experimentally. Thus, it is plausible to think that low
virulent viral strains (reduced cell killing ability, lack of colonization
skills), which possess a reproductive advantage within coinfected cells
(competition skills), will be naturally selected in a cell culture infection
with multiple viral strains. In the same line of thought, a stronger
relationship between both types of skills is conceivable, namely, a
trade-off between virulence (colonization) and reproductive advantage within
coinfected cells (competition).

In the present article, we make a first step towards transferring the
results obtained \emph{in vitro} (\citet{Ojosnegros2010}) to the \emph{in
vivo} situation under the assumption of a competition-colonization
trade-off. We extend the basic model of competition-colonization dynamics of
two different viral strains in cell culture by replacing the finite cell
monolayer with a constantly replenished pool of uninfected cells.
Furthermore, to account for the competition-colonization trade-off
postulated, we model the intracellular fitness of the viruses during
coinfection as being inversely proportional to their respective virulence.
In this relationship we archetypically capture the trade-off assumption.

We present a rigorous analytical study of this model and demonstrate that it
allows for local asymptotically stable coexistence of competitors and
colonizers. Moreover, the less virulent competitor is shown to have a
reproductive advantage reflected by its higher abundance and a higher number
of cells infected with it at equilibrium. This counterintuitive result is in
contrast to traditional modeling approaches of viral competition, in which
the equilibrium state is dominated by the most virulent strains (%
\citet{CoinfectionAndEvolutionVir}).

A straightforward generalization of this model is to consider more than two
viral variants. If we assume that different viral strains can only be
distinguished through their virulences, the question arises as to how a
distribution of virulences at the onset of infection is modified in the
course of the infection by the competition-colonization dynamics. In this
sense, the model describes the time evolution of a \textit{discrete}
distribution of virulences. While simulation results for a finite number of
viral strains will be presented elsewhere (Ojosnegros et al., in
preparation), herein we account for the very high phenotypic diversity of
viral populations by considering the continuum limit of this multiple viral
strains model. In this continuum limit, we consider a continuous spectrum of
virulence values and associate a different viral strain with each virulence
value. Thus, the time evolution of a \textit{continuous} distribution of
virulence is modeled. For this model's equations, we provide a proof of the
existence of solutions and present numerical approximations of solutions for
different initial distributions.

Our simulations suggest that initial continuous distributions of virulence
evolve towards a stationary distribution that is extremely skewed in favor
of the competitors. Thus, the model predicts attenuation of the virus
population and it might explain previous observations of suppression of
high-fitness mutants in various viral systems (%
\citet{Torre1990,Novella2004,Turner1999b,Bull2006}).

This article is organized as follows. In Section~\ref{sec:two-virus} we
formulate the basic model of two competing viral strains and illustrate our
model assumptions. In the Results Section a detailed analytical study of its
equilibrium behavior is presented. In Subsection~\ref{sec:multiple-virus},
the multiple-viral-strains model is introduced and in Subsection~\ref%
{sec:continuous}, we derive its continuum limit. The continuous-virulence
model is analyzed both analytically and numerically. We close in Section~\ref%
{sec:discussion} with discussing some of the model assumptions and
consequences for the evolution of virulence.

\section{Formulation of the two-viral-strains model}

\label{sec:two-virus}

Our modeling approach is based on the basic SIR model of virus dynamics (%
\citet{NowakMay},\citet{Perelson1999}), which we extend in order to model
two different viral populations infecting constantly replenished tissue. We
model singly infected and superinfected (doubly infected or coinfected)
cells. Thus, the time evolution of the concentration of two competing viral
strains, uninfected and infected cells is described by the following system
of ordinary differential equations 
\begin{eqnarray}
\dot{x} &=&\lambda -dx-\beta x(v_{1}+v_{2})  \notag \\
\dot{y}_{1} &=&\beta xv_{1}-\beta y_{1}v_{2}-a_{1}y_{1}  \notag \\
\dot{y}_{2} &=&\beta xv_{2}-\beta y_{2}v_{1}-a_{2}y_{2}
\label{Eq.2Virs.Model} \\
\dot{y}_{12} &=&\beta (y_{1}v_{2}+y_{2}v_{1})-\min (a_{1},a_{2})y_{12} 
\notag \\
\dot{v}_{1} &=&Ka_{1}y_{1}+cK\min (a_{1},a_{2})y_{12}-uv_{1}  \notag \\
\dot{v}_{2} &=&Ka_{2}y_{2}+(1-c)K\min (a_{1},a_{2})y_{12}-uv_{2}  \notag
\end{eqnarray}%
The variable $x$ models the concentration of uninfected cells with an
external constant supply of new cells at rate $\lambda $, dying at a rate $d$
and being infected with efficiency $\beta $. The variable $v_{1}$ resp. $%
v_{2}$ describes the concentration of strain 1 resp. strain 2. The variable $%
y_{i}$ represents the concentration of cells infected solely with strain $i.$
These cells die and release viral offspring at rate $a_{i},$ the virulence
of strain $i.$ The variable $y_{12}$ models the concentration of cells
infected with both viral strains. These cells die and release viral
offspring at rate $\min (a_{1},a_{2})$ (more on this below). Free virus of
type $i$ is produced at rate $k_{i}=Ka_{i}$, where $K$ is the burst size,
and inactivated at rate $u$. The parameter $c$ denotes the proportion of
strain $1$ produced at the burst of coinfected cells $y_{12}.$ The state of
the system at time $t$ is denoted $S(t)=(x(t),\,y_{1}(t),\,y_{2}(t),%
\,y_{12}(t),\,v_{1}(t),\,v_{2}(t))^{T}$.

We make the following general assumptions about the parameters:

\begin{itemize}
\item All parameters are positive.

\item The efficiency with which strain 1 resp. strain 2 infects uninfected
cells or singly infected cells ($y_{1}$ or $y_{2}$) is equal and denoted by $%
\beta .$

\item The death rates of strain 1 resp. strain 2 are equal and denoted by $u$%
.
\end{itemize}

Furthermore, based on the experimental results presented in %
\citet{Ojosnegros2010}, we assume:

\begin{itemize}
\item The burst size of singly and coinfected cells is equal and denoted by $%
K$.

\item The death rate of coinfected cells is equal to the death rate of cells
singly infected with the least virulent virus, $a_{12}=\min (a_{1},\,a_{2})$%
. In other words, this rate is imposed by the least virulent viral strain.

\item To account for the competition-colonization trade-off postulated, we
model the intracellular fitness of the viruses during coinfection as being
inversely proportional to their respective virulence by setting $%
c:=a_{1}^{-1}/(a_{1}^{-1}+a_{2}^{-1}).$ In this relationship we
archetypically capture the assumption that, viruses compensate the lack of
colonization skills with intracellular competition abilities and vice versa.
\end{itemize}

Without loss of generality, we can assume $a_{1}\leq a_{2}.$ Thus, strain 1
is the competitor and strain 2 is the colonizer and the last three equations
of (\ref{Eq.2Virs.Model}) become%
\begin{eqnarray*}
\dot{y}_{12} &=&\beta y_{1}v_{2}+\beta y_{2}v_{1}-a_{1}y_{12} \\
\dot{v}_{1} &=&Ka_{1}y_{1}+cKa_{1}y_{12}-uv_{1} \\
\dot{v}_{2} &=&Ka_{2}y_{2}+(1-c)Ka_{1}y_{12}-uv_{2}
\end{eqnarray*}%
In \citet{Ojosnegros2010}, the model (\ref{Eq.2Virs.Model}) with $\lambda =0$%
, $d=0,$ $a_{1}<a_{2}$ and an unconstrained (i.e. independent of $a_{1}$)
parameter $c>1/2,$ was introduced to describe two competing viral strains in
cell culture. These special \emph{in vitro} conditions of a fixed and
limited amount of target cells allowed for an analytical treatment of the
system in the large initial virus load limit which is not possible in the
more general case of system (\ref{Eq.2Virs.Model}) considered here.

\section{Results}

\subsection{Establishing infection}

To identify the conditions on the parameters of the model that imply spread
of at least one of the viral strains, we analyze the stability of the
(obvious) equilibrium point%
\begin{equation*}
S^{(0)}=(x^{(0)},y_{1}^{(0)},y_{2}^{(0)},y_{12}^{(0)},v_{1}^{(0)},v_{2}^{(0)})^{T}:=(\lambda /d,0,0,0,0,0)^{T}
\end{equation*}%
at which the infection dies out. The Jacobian matrix $J$ of system (\ref%
{Eq.2Virs.Model}) evaluated at $S^{(0)}$ is%
\begin{equation*}
J(S^{(0)})=%
\begin{pmatrix}
-d & 0 & 0 & 0 & -\beta \lambda /d & -\beta \lambda /d \\ 
0 & -a_{1} & 0 & 0 & \beta \lambda /d & 0 \\ 
0 & 0 & -a_{2} & 0 & 0 & \beta \lambda /d \\ 
0 & 0 & 0 & -a_{1} & 0 & 0 \\ 
0 & Ka_{1} & 0 & cKa_{1} & -u & 0 \\ 
0 & 0 & Ka_{2} & (1-c)Ka_{1} & 0 & -u%
\end{pmatrix}%
\end{equation*}%
The eigenvalues of this matrix are%
\begin{equation*}
-d,\quad -a_{1},\quad \frac{-d(a_{i}+u)+\Delta _{i}}{2d},\quad \frac{%
-d(a_{i}+u)-\Delta _{i}}{2d},\qquad i=1,2,
\end{equation*}%
where $\Delta _{i}:=\sqrt{d^{2}(a_{i}-u)^{2}+4Ka_{i}\beta \lambda d}$ for $%
i=1,2.$ All\ eigenvalues are real given that all parameters are assumed to
be positive. This equilibrium becomes unstable as soon as at least one
eigenvalue is positive. This happens if and only if $-d(a_{1}+u)+\Delta
_{1}>0$ or $-d(a_{2}+u)+\Delta _{2}>0$ which is equivalent to $%
d(a_{1}-u)^{2}+4Ka_{1}\beta \lambda >d(a_{1}+u)^{2}$ or $%
d(a_{2}-u)^{2}+4Ka_{2}\beta \lambda >d(a_{2}+u)^{2}.$ The latter expression
is in turn equivalent to $K\beta \lambda >du.$ In other words, it is
sufficient for viral spread that%
\begin{equation}
R_{0}:=\frac{K\beta \lambda }{du}>1  \label{eq.SpreadCondition}
\end{equation}%
For generic parameter values (the fine tuning $K\beta \lambda =du$ cannot be
expected), this condition is also necessary, because $K\beta \lambda <du$
implies that $S^{(0)}$ is asymptotically stable.

If we consider initial conditions in which $y_{i}(0)=0,$ $y_{12}(0)=0,$ $%
v_{i}(0)=0$ and $v_{j}(0)\neq 0,$ where $i,j\in \{1,2\},i\neq j,$ the model
reduces to a simple SIR model of single viral infection (\citet{Bonhoeffer}, %
\citet{NowakMay}, see also \citet{Korobeinikov} for global results) and we
recognize the magnitude $R_{0}$ as the well known \emph{basic reproductive
number} of the infection system. Condition (\ref{eq.SpreadCondition}) can
also be expressed as $M:=K\beta \lambda -du>0.$ The magnitude $M$ turns out
to be algebraically very helpful for our further analysis of the model.

Having determined a condition on the parameter values that characterizes the
event of viral spread, we next ask whether under these circumstances, the
system admits a steady state in which both viral strains can coexist. The
opposite steady state scenario would be that one of the viral strains
outcompetes the other. To this end, we examine further fixed points of the
system and their stability.

\subsection{Further fixed points}

Performing algebraic manipulations we found the following non-trivial fixed
points of the system (\ref{Eq.2Virs.Model}):%
\begin{equation*}
S^{(1)}=%
\begin{pmatrix}
x^{(1)} \\ 
y_{1}^{(1)} \\ 
y_{2}^{(1)} \\ 
y_{12}^{(1)} \\ 
v_{1}^{(1)} \\ 
v_{2}^{(1)}%
\end{pmatrix}%
:=\frac{1}{\beta }%
\begin{pmatrix}
u/K \\ 
M/(a_{1}K) \\ 
0 \\ 
0 \\ 
M/u \\ 
0%
\end{pmatrix}%
\end{equation*}

\begin{equation*}
S^{(2)}=%
\begin{pmatrix}
x^{(2)} \\ 
y_{1}^{(2)} \\ 
y_{2}^{(2)} \\ 
y_{12}^{(2)} \\ 
v_{1}^{(2)} \\ 
v_{2}^{(2)}%
\end{pmatrix}%
:=\frac{1}{\beta }%
\begin{pmatrix}
u/K \\ 
0 \\ 
M/(a_{2}K) \\ 
0 \\ 
0 \\ 
M/u%
\end{pmatrix}%
\end{equation*}

\begin{equation*}
S^{\ast }=%
\begin{pmatrix}
x^{\ast } \\ 
y_{1}^{\ast } \\ 
y_{2}^{\ast } \\ 
y_{12}^{\ast } \\ 
v_{1}^{\ast } \\ 
v_{2}^{\ast }%
\end{pmatrix}%
:=\frac{1}{\beta K}%
\begin{pmatrix}
u \\ 
a_{2}uM/(a_{1}(M+u(a_{1}+a_{2}))) \\ 
a_{1}uM/(a_{2}(M+u(a_{1}+a_{2}))) \\ 
M^{2}/(a_{1}(M+u(a_{1}+a_{2}))) \\ 
a_{2}KM/(u(a_{1}+a_{2})) \\ 
a_{1}KM/(u(a_{1}+a_{2}))%
\end{pmatrix}%
\end{equation*}%
and%
\begin{equation*}
S^{-}=%
\begin{pmatrix}
x^{-} \\ 
y_{1}^{-} \\ 
y_{2}^{-} \\ 
y_{12}^{-} \\ 
v_{1}^{-} \\ 
v_{2}^{-}%
\end{pmatrix}%
:=\frac{1}{\beta K}%
\begin{pmatrix}
\beta K\lambda /d \\ 
(a_{2}M-ud\zeta )/((c-1)d(a_{1}+a_{2})) \\ 
-(a_{1}M-ud\zeta )/(cd(a_{1}+a_{2})) \\ 
-(a_{1}a_{2}M+ud\zeta ((a_{1}+a_{2})c-a_{2}))/(c(c-1)da_{1}(a_{1}+a_{2})) \\ 
\zeta K \\ 
-\zeta K%
\end{pmatrix}%
\end{equation*}%
where $\zeta $ is a root of the polynomial $duZ^{2}+(-c\beta Ka_{2}\lambda
-\beta K\lambda a_{1}c+\beta K\lambda a_{1}+da_{2}u-a_{1}ud)Z$ $%
-a_{1}a_{2}ud+\beta K\lambda a_{1}a_{2}$ in the indeterminate $Z.$ Since we
are only interested in real, non-negative solutions, the fixed point $S^{-}$
can be disregarded, because it has at least one negative component for
whatever real value $\zeta $ might take (including $\zeta =0,$ which implies 
$y_{2}^{-}=-a_{1}M/(c\beta Kd(a_{1}+a_{2}))$). By the assumed positivity of
the parameters and by $M>0$ (we assume that viral spread is possible) all
other fixed points are (component-wise) non-negative and thus biologically
meaningful. Summarizing, we have two equilibria, $S^{(1)}$ and $S^{(2)},$ in
which one of the viral strains outcompetes the other, and one coexistence
equilibrium $S^{\ast }$. The following subsections are devoted to studying
their properties.

\subsection{The coexistence equilibrium}

The Jacobian matrix $J(S^{\ast })$ of system (\ref{Eq.2Virs.Model})
evaluated at the equilibrium point $S^{\ast }$ equals%
\begin{equation*}
\begin{pmatrix}
-\beta \lambda K/u & 0 & 0 & 0 & -u/K & -u/K \\ 
\frac{a_{2}M}{u(a_{1}+a_{2})} & -\frac{a_{1}M}{u(a_{1}+a_{2})}-a_{1} & 0 & 0
& u/K & -\frac{a_{2}uM}{a_{1}K(M+u(a_{1}+a_{2}))} \\ 
\frac{a_{1}M}{u(a_{1}+a_{2})} & 0 & -\frac{a_{2}(M+u(a_{1}+a_{2}))}{%
u(a_{1}+a_{2})} & 0 & -\frac{a_{1}uM}{a_{2}K(M+u(a_{1}+a_{2}))} & u/K \\ 
0 & \frac{a_{1}M}{u(a_{1}+a_{2})} & \frac{a_{2}M}{u(a_{1}+a_{2})} & -a_{1} & 
\frac{a_{1}uM}{a_{2}K(M+u(a_{1}+a_{2}))} & \frac{a_{2}uM}{%
a_{1}K(M+u(a_{1}+a_{2}))} \\ 
0 & Ka_{1} & 0 & \frac{Ka_{1}a_{2}}{(a_{1}+a_{2})} & -u & 0 \\ 
0 & 0 & Ka_{2} & \frac{Ka_{1}^{2}}{(a_{1}+a_{2})} & 0 & -u%
\end{pmatrix}%
\end{equation*}%
The characteristic polynomial of this matrix is too long to be displayed.
Thus, we performed its analysis using computer algebra and symbolic
computation. Under the premise of viral spread, i.e. $M>0,$ we used the
computer algebra system Maple$^{\text{{\small TM}}}$ to show that all
coefficients are positive. Furthermore, we used Maple$^{\text{{\small TM}}}$
to construct Routh's table (\citet{OnRouthsCriterion}) and verified that all
entries in its first column are positive (see supplementary material). By
Routh's criterion (\citet{OnRouthsCriterion}), all roots of the
characteristic polynomial must lie strictly to the left of the imaginary
axes. As a consequence, the equilibrium point $S^{\ast }$ in which both
viral strains can coexist is a local asymptotically stable fixed point. This
holds for all positive parameter values, provided $M>0.$

The peculiarity of this coexistence equilibrium is that the viral load at
equilibrium of the competitor, $v_{1}^{\ast }=a_{2}M/(\beta u(a_{1}+a_{2}))$%
, is proportional to the \textit{relative} virulence of the colonizer.
Similarly, the viral load at equilibrium of the colonizer, $v_{2}^{\ast
}=a_{1}M/(\beta u(a_{1}+a_{2}))$, is proportional to the relative virulence
of the competitor. Thus, the two-viral-strains system (\ref{Eq.2Virs.Model})
not only allows for coexistence of both viral strains, but it confers the
less virulent competitor a reproductive advantage over the more virulent
colonizer. This discrepancy is reflected in the higher concentration of
competitors, $v_{1}^{\ast }/v_{2}^{\ast }=a_{2}/a_{1}>1$, and the higher
concentration of cells infected with competitors, $y_{1}^{\ast }/y_{2}^{\ast
}=(a_{2}/a_{1})^{2}>1$, at equilibrium. The property that makes this
phenomenon possible is the inverse proportionality between virulence and
intracellular fitness during coinfection, as explained in more detail in the
Discussion.

\subsection{Single viral strain equilibria}

The existence of a local asymptotically stable coexistence equilibrium
suggests that a viral strain can bear the presence of another one. However,
to fully address the question as to whether system (\ref{Eq.2Virs.Model})
always allows for a second viral strain to invade tissue already infected
with a different strain, we examine the stability of the equilibria $S^{(1)}$
and $S^{(2)}$, in which one of the viral strains extrudes the other.

The Jacobian matrix of system (\ref{Eq.2Virs.Model}) evaluated at the point $%
S^{(1)}$ is%
\begin{equation*}
J(S^{(1)})=%
\begin{pmatrix}
-M/u-d & 0 & 0 & 0 & -u/K & -u/K \\ 
M/u & -a_{1} & 0 & 0 & u/K & -M/(a_{1}K) \\ 
0 & 0 & -\frac{M}{u}-a_{2} & 0 & 0 & u/K \\ 
0 & 0 & M/u & -a_{1} & 0 & M/(a_{1}K) \\ 
0 & Ka_{1} & 0 & Ka_{1}a_{2}/(a_{1}+a_{2}) & -u & 0 \\ 
0 & 0 & Ka_{2} & Ka_{1}^{2}/(a_{1}+a_{2}) & 0 & -u%
\end{pmatrix}%
\end{equation*}%
Assuming viral spread, ($M>0$), we showed that the leading coefficient of
the characteristic polynomial of $J(S^{(1)})$ is positive, whereas the
independent coefficient is negative (see supplementary material). By Routh's
criterion, at least one root of the characteristic polynomial has positive
real part. Therefore, the equilibrium point $S^{(1)}$ in which the
competitor outcompetes the colonizer is not stable. Analogously, we found
that the equilibrium point $S^{(2)}$ in which the colonizer outcompetes the
competitor is also unstable. Both statements hold true for all positive
parameter values, provided that $M>0$.

However, it is worth mentioning that any trajectory $S(t)$, for which one
viral strain, say of type $i,$ and all cells infected or coinfected with it
have disappeared at some point in time $s$, would stay confined in the
corresponding hyperplane $H_{i}:=\{(x,y_{1},y_{2},y_{12},v_{1},v_{2})^{T}%
\mid y_{i}=y_{12}=v_{i}=0\}$ for all $t\geq s$. As mentioned in Subsection
2, \emph{within} the corresponding hyperplanes, the equilibria $S^{(1)}$ and 
$S^{(2)}$ become asymptotically stable, provided $R_{0}>1$. Whether a
trajectory starting outside $H_{i}$ flows into the hyperplane or not remains
to be analytically studied. Our simulations do not show any evidence for
this type of behavior.

\subsection{Multiple-viral-strains model}

\label{sec:multiple-virus}

A straightforward generalization of our model (\ref{Eq.2Virs.Model}) that
accounts for the experimentally observed diversity of viral populations is
to consider more than two competing viral strains. This generalization
raises the question of how many viruses can coinfect a cell simultaneously.
For example, the multiplicity of HIV-infected spleen cells has been reported
between 1 and 8 with mean 3.2 (\citet{CoinfectionSize}). However, for the
sake of mathematical simplicity, we consider here the case in which at most
two viruses can coinfect a cell. We assume that we can distinguish each of
the viral strains via their corresponding virulences $a_{i}$. As above, we
assume inverse proportionality between virulence and intracellular fitness
during coinfection. In other words, the proportion of strain $i$ produced at
the burst of cells $y_{ij}$ coinfected with $v_{i}$ and $v_{j}$ is given by $%
c_{i}:=a_{i}^{-1}/\left( a_{i}^{-1}+a_{j}^{-1}\right) ,$ for all $i,j\in
\{1,...,n\},$ $i<j$ , where $n\in 
\mathbb{N}
$ is the total number of viral strains modeled. Thus, the equations for the
generalized model read:%
\begin{eqnarray*}
\dot{x} &=&\lambda -dx-\beta x\sum\limits_{j=1}^{n}v_{j} \\
\dot{y}_{i} &=&\beta xv_{i}-\beta y_{i}\left( \sum\limits_{\substack{ j=1 
\\ j\neq i}}^{n}v_{j}\right) -a_{i}y_{i},\text{ \ \ }i=1,...,n \\
\dot{y}_{lj} &=&\beta (y_{l}v_{j}+y_{j}v_{l})-\min (a_{l},a_{j})y_{lj},\text{
\ \ }l,j=1,...,n\text{ and }l<j \\
\dot{v}_{i} &=&Ka_{i}y_{i}+a_{i}^{-1}K\left( \sum\limits_{\substack{ l,j  \\ %
l<j}}\frac{1}{a_{l}^{-1}+a_{j}^{-1}}w_{i}(l,j)\min
(a_{l},a_{j})y_{lj}\right) -uv_{i},\text{ \ \ }i=1,...,n
\end{eqnarray*}%
where $w_{i}(l,j)=1$ if $l=i$ or $j=i,$ and otherwise $w_{i}(l,j)=0.$ The
model does not explicitly account for the order of infection. Nevertheless,
to increase the symmetry of the model and to simplify the notation, we will
consider the order of infection events and separately model the populations $%
y_{lj}$ and $y_{jl}$, where the order of the indices indicates the order of
infection with the viral strains $v_{l}$ and $v_{j}$. With this notation, in
general, $y_{lj}\not=y_{jl}$, and the variable $y_{12}$ in the
two-viral-strains model (\ref{Eq.2Virs.Model}) refers to $y_{12}+y_{21}$. To
be consistent, we have to make sure that for $l\neq j,$ the magnitude $%
y_{lj}+y_{jl}$ obeys the corresponding equation, that is%
\begin{equation*}
\dot{y}_{lj}+\dot{y}_{jl}=\beta (y_{l}v_{j}+y_{j}v_{l})-\min
(a_{l},a_{j})(y_{lj}+y_{jl})
\end{equation*}%
To ensure this, the equation for $y_{lj}$ becomes%
\begin{equation*}
\dot{y}_{lj}=\beta y_{l}v_{j}-\min (a_{l},a_{j})y_{lj},\text{ \ \ }%
l,j=1,...,n\text{ s.t. }l\neq j
\end{equation*}%
Summarizing, we obtain the following model%
\begin{eqnarray}
\dot{x} &=&\lambda -dx-\beta x\sum\limits_{j=1}^{n}v_{j}  \notag \\
\dot{y}_{i} &=&\beta xv_{i}-\beta y_{i}\left( \sum\limits_{\substack{ j=1 
\\ j\neq i}}^{n}v_{j}\right) -a_{i}y_{i},\text{ \ \ }i=1,...,n  \notag \\
\dot{y}_{lj} &=&\beta y_{l}v_{j}-\min (a_{l},a_{j})y_{lj},\text{ \ \ }%
l,j=1,...,n\text{ s.t. }l\neq j  \label{Eq.nVirs.Model} \\
\dot{v}_{i} &=&Ka_{i}y_{i}+a_{i}^{-1}\left( \sum\limits_{\substack{ j=1  \\ %
j\neq i}}^{n}\frac{1}{a_{i}^{-1}+a_{j}^{-1}}K\min
(a_{i},a_{j})(y_{ij}+y_{ji})\right) -uv_{i},\text{ \ \ }i=1,...,n  \notag
\end{eqnarray}%
Note that the magnitude $\sum\limits_{j=1}^{n}v_{j}(t)$ is the total viral
population at any given point in time $t$.

Since the number of equations in this model grows quadratically with the
number $n$ of viral strains, it becomes rather involved to analyze it. In
(Ojosnegros et al., in preparation) the results of numerical simulation for
numerically tractable values of $n$ are presented. Here, in compliance with
the quasispecies view of viral populations, we devise a new approach to
studying the evolution of virulence and consider the continuum limit of the
multistrain model (\ref{Eq.nVirs.Model}). In this continuum limit, we
consider a continuous spectrum of virulence values and identify viral
strains with virulence values. We call the resulting continuum limit the
continuous-virulence model. In this model the viral quasispecies is
represented by a time-dependent continuous distribution of virulence. Unlike
the discrete multiple-viral-strains model, the continuum approach allows us
to study the virulence distribution of diverse RNA virus populations in a
manner independent of the number of different strain types.

\subsection{Continuous-virulence model}

\label{sec:continuous}

We identify viral strains with their virulence $a$ and denote by $v(a,t)$
the density of viruses of type~$a$ at time $t$. If we consider an interval $%
[a_{1},a_{2}]\subset (0,1)$ of possible virulences, then the initial
distribution of virulences is defined by a continuos density function $%
v(\cdot ,0):%
\mathbb{R}
\rightarrow 
\mathbb{R}
$ which vanishes outside the interval $[a_{1},a_{2}].$ The continuum limit
of model (\ref{Eq.nVirs.Model}) is therefore the following initial value
problem (also known as Cauchy problem):%
\begin{eqnarray}
\dot{x}(t) &=&\lambda -dx(t)-\beta x(t)\int\limits_{a_{1}}^{a_{2}}v(\xi
,t)d\xi  \notag \\
\frac{\partial y}{\partial t}(a,t) &=&\beta x(t)v(a,t)-\beta y(a,t)\left(
\int\limits_{a_{1}}^{a_{2}}v(\xi ,t)d\xi \right) -ay(a,t)  \notag \\
\frac{\partial z}{\partial t}(a,b,t) &=&\beta y(a,t)v(b,t)-\min (a,b)z(a,b,t)
\label{Eq.Cont.Model} \\
\frac{\partial v}{\partial t}(a,t) &=&Kay(a,t)+a^{-1}K\left(
\int\limits_{a_{1}}^{a_{2}}\frac{1}{a^{-1}+b^{-1}}\min
(a,b)(z(a,b,t)+z(b,a,t))db\right) -uv(a,t)  \notag \\
x(0) &=&x_{0},\text{ }y(\xi ,0)=y_{0}(\xi ),\text{ }z(\vartheta ,\mu
,0)=z_{0}(\vartheta ,\mu ),\text{ }v(\xi ,0)=v_{0}(\xi )  \notag
\end{eqnarray}%
For every $t\in \mathbb{R}_{\geq 0},$ the function $z(\cdot ,\cdot
,t):[a_{1},a_{2}]\times \lbrack a_{1},a_{2}]\rightarrow 
\mathbb{R}
$ describes the density of coinfected cells with respect to the
two-dimensional (Lebesgue) measure on $%
\mathbb{R}
^{2}$. The value $z(a,b,t)$ is only meaningful for our modeling purposes
outside the diagonal $a=b.$ Note that the exception $j\neq i$ in the sums of
the original discrete model can be neglected here because the values of a
real valued function on a set of measure zero do not modify the value of the
integral. If we assume that a solution of this initial value problem exists,
then we can derive the following expressions: Setting $V(t):=%
\int_{a_{1}}^{a_{2}}v(\xi ,t)d\xi $ we have, for the first equation, $\dot{x}%
(t)=\lambda -dx-\beta x(t)V(t)$ and thus%
\begin{equation}
x(t)=\left( x_{0}+\int\limits_{0}^{t}\lambda e^{W(\xi )}d\xi \right)
e^{-W(t)}  \label{Eq.Exp.ForX}
\end{equation}%
where $W(\theta ):=\int\nolimits_{0}^{\theta }d+\beta V(\tau )d\tau .$ The
second equation can be solved as%
\begin{equation*}
y(a,t)=\left( y_{0}(a)+\beta \int\limits_{0}^{t}x(\tau )v(a,\tau
)e^{U_{a}(\tau )}d\tau \right) e^{-U_{a}(t)}
\end{equation*}%
where $U_{a}(\theta ):=\int\nolimits_{0}^{\theta }a+\beta V(\tau )d\tau .$
Substituting the expression for $x(t)$ into the expression for $y(a,t)$ gives%
\begin{eqnarray}
y(a,t) &=&\left( y_{0}(a)+\beta \int\limits_{0}^{t}v(a,\tau )\left(
x_{0}+\int\limits_{0}^{\tau }\lambda e^{W(\xi )}d\xi \right) e^{U_{a}(\tau
)-W(\tau )}d\tau \right) e^{-U_{a}(t)}  \notag \\
&=&\left( y_{0}(a)+\beta \int\limits_{0}^{t}v(a,\tau )\left(
x_{0}+\int\limits_{0}^{\tau }\lambda e^{W(\xi )}d\xi \right) e^{(a-d)\tau
}d\tau \right) e^{-U_{a}(t)}  \label{Eq.Exp.ForY}
\end{eqnarray}%
The third equation yields%
\begin{equation}
z(a,b,t)=\left( z_{0}(a,b)+\beta \int\limits_{0}^{t}y(a,\eta )v(b,\eta
)e^{\min (a,b)\eta }d\eta \right) e^{-\min (a,b)t}  \label{Eq.Exp.ForZ}
\end{equation}%
Substituting the expression for $y(a,t)$ allows us to express $%
z(a,b,t)+z(b,a,t)$ in terms of $v(a,\tau ),$ $v(b,\tau ),$ and integrals
involving them as%
\begin{equation*}
z(a,b,t)+z(b,a,t)=\left( z_{0}(a,b)+z_{0}(b,a)+R(a,b,t)\right) e^{-\min
(a,b)t}
\end{equation*}%
where%
\begin{eqnarray*}
R(a,b,t):= &&\beta \int\limits_{0}^{t}\left( \left( v(b,\eta
)y_{0}(a)+v(a,\eta )y_{0}(b)+Q(a,b,\eta )\right) e^{\min (a,b)\eta
-U_{a}(\eta )}\right) d\eta \\
Q(a,b,\eta ):= &&\beta \int\limits_{0}^{\eta }\left( v(b,\eta )v(a,\tau
)e^{(a-d)\tau }+v(a,\eta )v(b,\tau )e^{(b-d)\tau }\right) \left(
x_{0}+\int\limits_{0}^{\tau }\lambda e^{W(\xi )}d\xi \right) d\tau
\end{eqnarray*}

Finally, we substitute the expressions obtained for $y(a,t)$ and $z(a,b,t)$
into the differential equation for $v(a,t)$ obtaining%
\begin{multline}
\frac{\partial v}{\partial t}(a,t) \,\, = \,\, Ka\left( \left(
y_{0}(a)+\int\limits_{0}^{t}\beta v(a,\tau )\left(
x_{0}+\int\limits_{0}^{\tau }\lambda e^{W(\xi )}d\xi \right) e^{(a-d)\tau
}d\tau \right) e^{-U_{a}(t)}\right) + \\
+ a^{-1}K\left( \int\limits_{a_{1}}^{a_{2}}\frac{1}{a^{-1}+b^{-1}}\min
(a,b)(z_{0}(a,b)+z_{0}(b,a)+R(a,b,t))e^{-\min (a,b)t}db\right) -uv(a,t)
\label{eq.IntegroPDEonv(a,t)}
\end{multline}%
A solution of the system (\ref{Eq.Cont.Model}) necessarily has to fulfill
this integro-partial differential equation for the function $v(a,t)$. On the
other hand, a solution of the latter equation that is continuous on $%
[a_{1},a_{2}]\times 
\mathbb{R}
$ and partially differentiable with respect to $t,$ and satisfies $v(\xi
,0)=v_{0}(\xi ),$ allows for constructing a solution of the system (\ref%
{Eq.Cont.Model}) by means of substitution of $v(a,t)$ into the expressions (%
\ref{Eq.Exp.ForX}), (\ref{Eq.Exp.ForY}), and (\ref{Eq.Exp.ForZ}). Based on
this idea, we show in Appendix~\ref{app:existence} the existence of
solutions of the system (\ref{Eq.Cont.Model}).

\subsubsection{Simulation results}

In order to explore the dynamics of the continuous-virulence model, we
numerically solved the Cauchy problem (\ref{Eq.Cont.Model}), as described in
more detail in Appendix~\ref{app:numerics}, using typical parameter values
given in Table~1. Because $y$, $z$, and $v$ represent concentration
densities, the units of $y$ and $v$ are [concentration]/[virulence] and the
unit of $z$ is [concentration]/[virulence]$^{2}$. Given that the unit of
virulence is [Time]$^{-1}$, $y$, $z$, and $v$ are measured in units of
[Time]/[Volume], [Time]$^{2}$/[Volume], and [Time]/[Volume], respectively.
The variable $x$ represents a concentration and its unit is therefore
[Volume]$^{-1}$. 
\begin{table}[hbt]
\begin{center}
\begin{tabular}{llll}
\hline
\textbf{Parameter} & \textbf{Description} & \textbf{Value} & \textbf{Units}
\\ \hline
$\lambda$ & Natural growth rate of uninfected population & $10^{5}$ & $%
(ml*h)^{-1}$ \\ 
$d$ & Natural death rate of uninfected population & $0.05$ & $h^{-1}$ \\ 
$\beta$ & Rate of infection & $5 \cdot 10^{-8}$ & $ml/h$ \\ 
$K$ & Burst size & 150 & Dimensionless \\ 
$u$ & Clearance rate of free virus & 0.15 & $h^{-1}$ \\ \hline
\end{tabular}%
\end{center}
\caption{ {\protect\footnotesize {Parameters of the model of evolution of
virulence during infection}}}
\end{table}

Starting from non infected tissue, i.e. $x_{0}=\lambda /d,$ $y_{0}(\xi
)\equiv 0$ and $z_{0}(\vartheta ,\mu )\equiv 0$, we studied three different
initial unnormalized continuous distributions of virulences given by the
densities $v(a,0)=v_{0}(a),$ $a\in \lbrack 0.01,0.5]:$

\begin{enumerate}
\item A uniform distribution defined by $v_{0}(a)=1$ if $a\in \lbrack
0.01,\,0.5]$ and $v_{0}(a)=0$ otherwise

\item A Gaussian distribution with mean value $\mu =1/4$ and standard
deviation $\sigma =1/40$ given by $v_{0}(a)=e^{-800(a-1/4)^{2}}$

\item A mixture of two Gaussian distributions with mean values $\mu
_{1}=1/6, $ $\mu _{2}=1/3$, equal standard deviations $\sigma _{1}=\sigma
_{2}=1/300$ and unequal weights $\lambda _{1}=3/10,$ $\lambda _{2}=7/10$
given by $v_{0}(a)=(3/7)e^{-4500(a-1/6)^{2}}+(7/10)e^{-4500(a-1/3)^{2}}.$
\end{enumerate}

The time evolution of a flat initial virulence density is shown in snapshots
in Figure 1. After a very short absorption phase, the density takes an
exponential shape in favor of the most virulent part of the interval which
is greatly amplified during the first eleven hours of simulation time. After
approximately 130 hours a recession is observed, which by $t=140$ $h$ has
already decreased the populations' densities by one order of magnitude.
After approximately 190 hours a qualitative change takes place and the
distribution starts loosing its exponential shape to become a non-symmetric
unimodal distribution with mode around virulence $=0.257$ at $t=240$ $h.$
This distribution starts traveling to the left, becomes narrower and more
symmetric. By $t=600$ $h$ the distribution has moved further to the left and
is now almost symmetric. 
\begin{figure}[h]
\centerline{\includegraphics[width=7.0in]{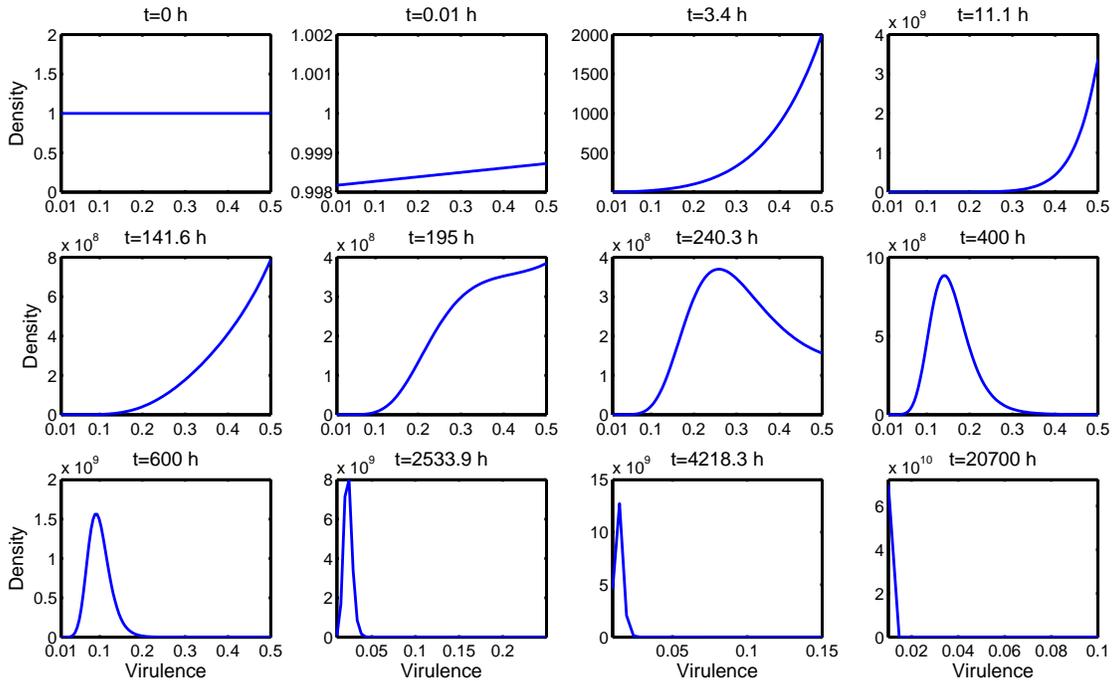}}
\caption{Time evolution of a uniform initial distribution of virulences.
Each panel shows the shape of the density function at the point in time
displayed in its titel.}
\label{Fig1}
\end{figure}
At this point, the dynamics become significantly slower and we start
observing an amplification effect. At $t=2534$ $h$ we observe a very narrow
Gaussian distribution approaching the left boundary of the interval $%
[0.01,0.5].$ Here the distribution starts changing its shape to become an
exponentially shaped distribution, this time in favor of low-virulence
competitors. The changes become very slow and, at least numerically, the
system seems to be reaching a stationary distribution, which is exponential
and highly in favor of the smallest virulence values. (See Appendix~\ref%
{app:figures2&3} for a similarly detailed description of Figures 2 and 3.) 
\begin{figure}[t]
\centerline{\includegraphics[width=7.0in]{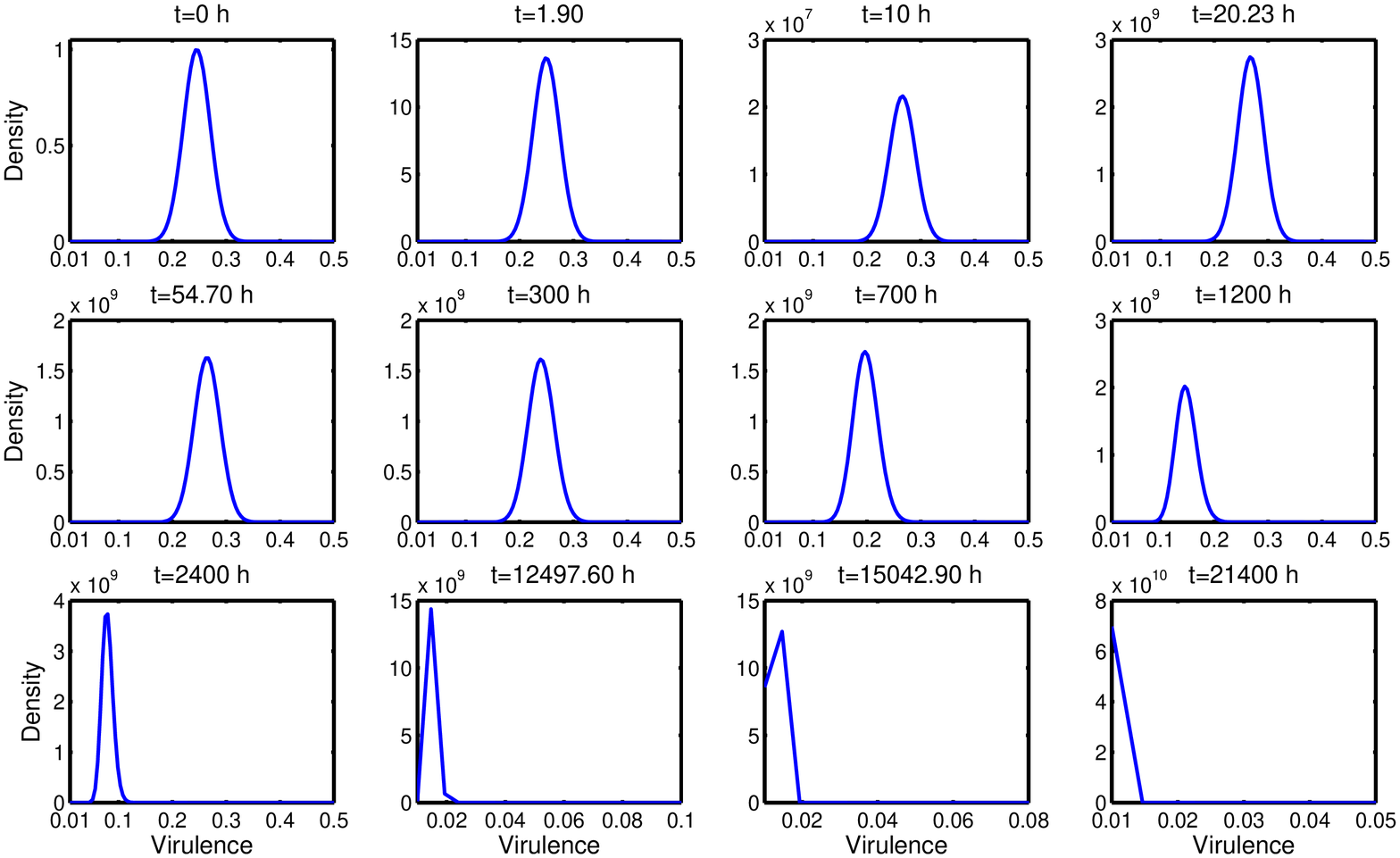}}
\caption{Time evolution of a Gaussian mixture initial distribution of
virulences.}
\label{Fig2}
\end{figure}

Despite the prominent differences between the initial distributions, Figures
2 and 3 reveal similar qualitative properties of the dynamics, namely, a
biphasic behavior comprising an initial phase in which the more virulent
parts of the density are amplified, followed by a second phase in which the
less virulent regions predominate. All three trajectories become very
similar once the density becomes unimodal, although the time scales are
significantly different, and all three reach very similar stationary
distributions. In the case of the uniform and the Gaussian distribution, the
stationary distributions reached differ only by small quantitative
discrepancies within the same order of magnitude.

For each of the three initial virulence densities $v_{0}(a)$, Figure~4 shows
the time evolution of the expected value of the virulence, $%
E(a)(t)=\int_{-\infty }^{\infty }bv(b,t)/\left\Vert v(b,t)\right\Vert db$,
which is obtained by normalizing at each point in time, $t$, as the system
of integro-partial differential equations is not norm-preserving. In this
graph, we can clearly identify the two different regimes of the biphasic
behavior described above. 
\begin{figure}[t]
\centerline{\includegraphics[width=7.0in]{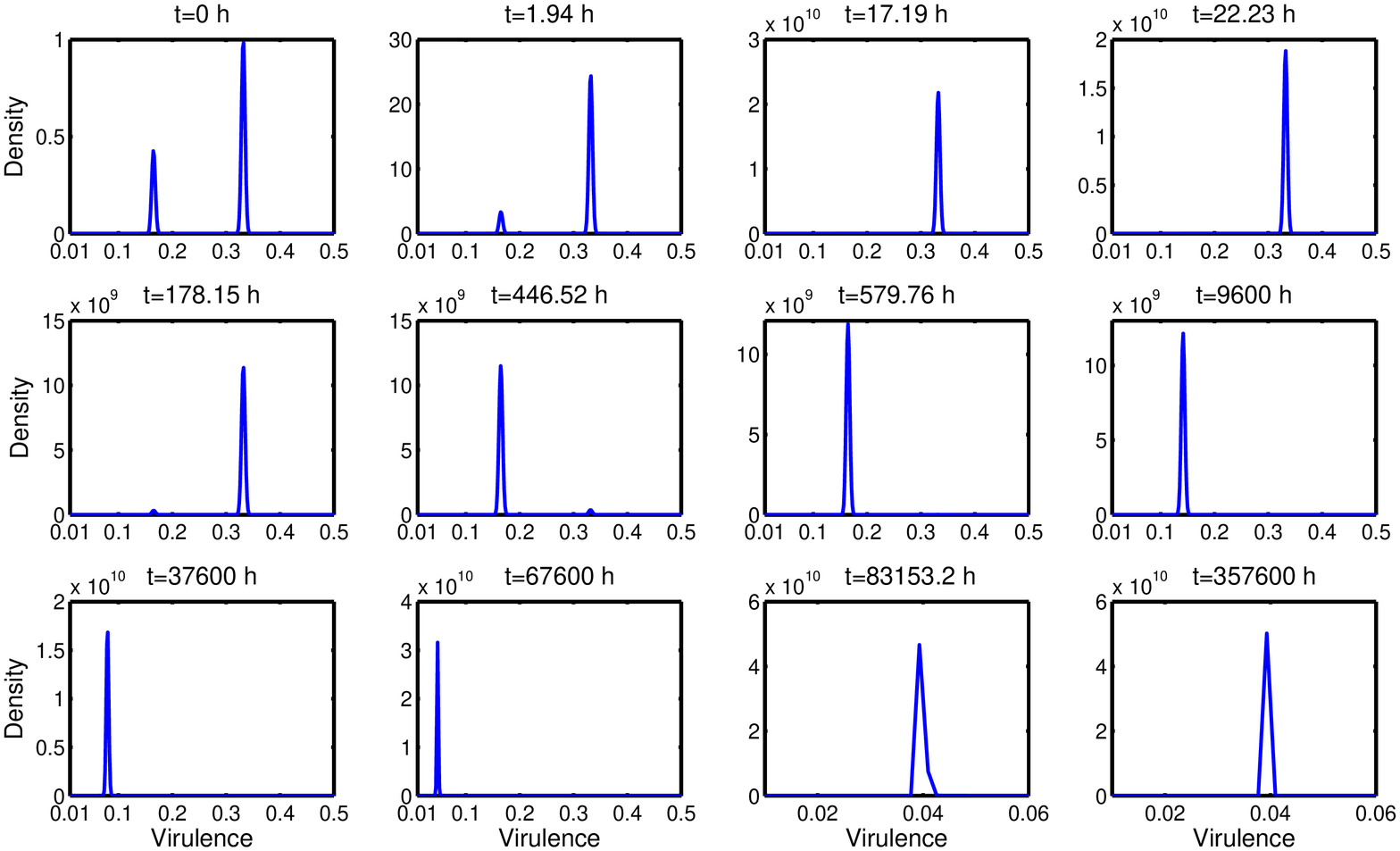}}
\caption{Time evolution of a Gaussian mixture initial distribution of
virulences.}
\label{Fig3}
\end{figure}
\begin{figure}[htb]
\centerline{\includegraphics[width=7.0in]{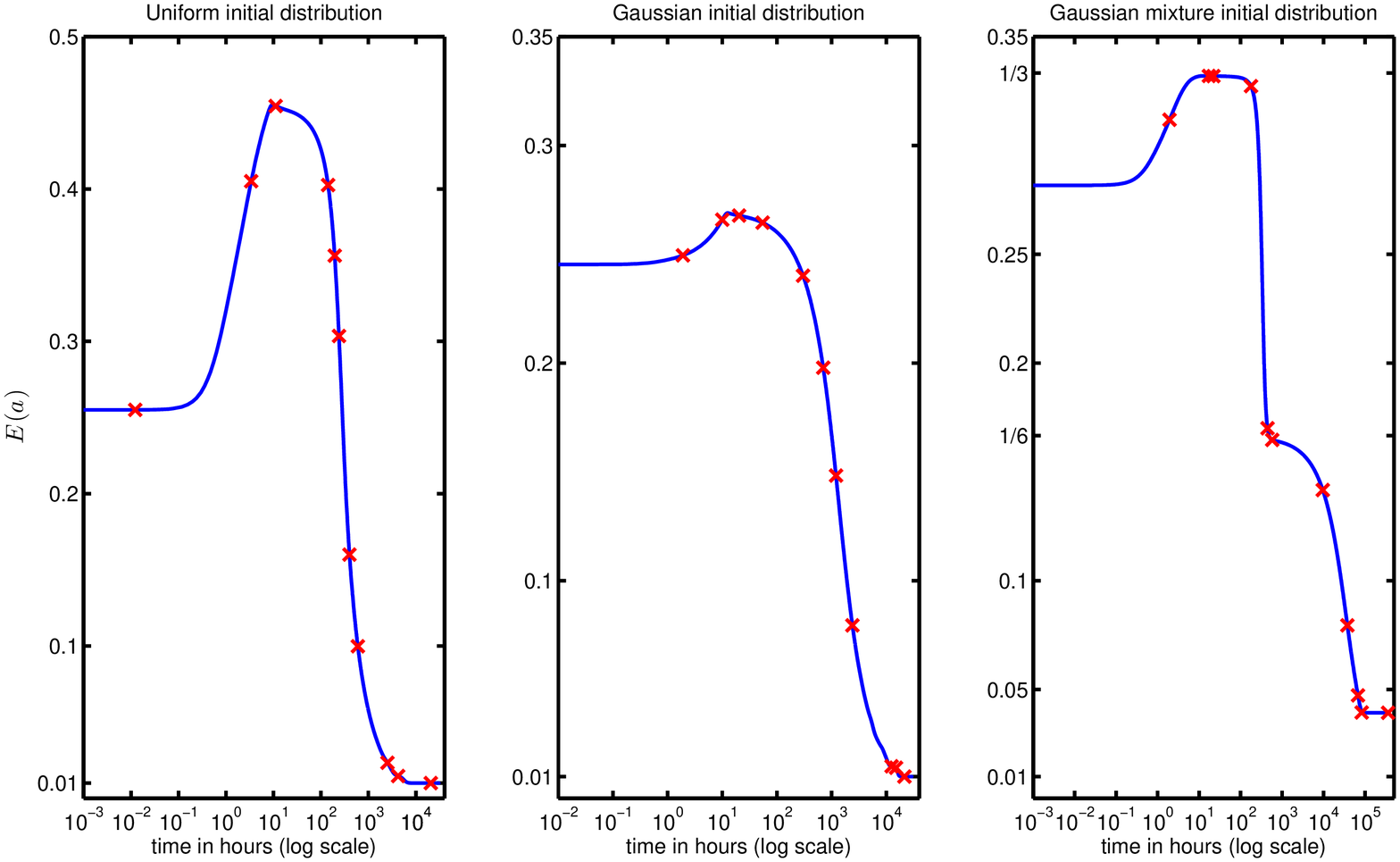}}
\caption{Time evolution of the expected value $E(a)=\protect\int_{-\infty
}^{\infty }bv(b,t)/\left\Vert v(b,t)\right\Vert db$ of virulence for three
different initial distributions. The crosses mark the value of $E(a)$ at the
time points ($>0$) depicted in the corresponding
evolution-of-distribution-figure (Figures 1-3).}
\label{Fig4}
\end{figure}

\section{Discussion}

\label{sec:discussion}

We have analyzed the evolution of virulence of an RNA viral quasispecies in
which the cell killing capacity (what we call the virulence $a_{i}$) of
viruses is inversely related to the intracellular viral fitness within
coinfected cells. In the case of two viral strains, this
competition-colonization trade-off allows for stable coexistence of
competitors and colonizers and each virus type can be invaded by the other,
whenever the conditions for viral spread are given. These conditions do not
depend on the particular virulence values $a_{i}.$ However, the population
levels at the coexistence equilibrium do depend on the particular virulence
values $a_{i},$ and, as we saw above, this dependency is in favor of the
least virulent viral strain.

Generalizing this two-viral-strains model to multiple viral strains is
conceptually straightforward, but the resulting system of differential
equations is difficult to study analytically. Moreover, the lack of accurate
experimental measurements of the actual number of (in terms of virulence)
different strains contained in a phenotypically diverse viral population
limits the applicability of this modeling approach. Furthermore, the
quadratic dependency of the number of equations on the number of strains $n$
(recall Subsection 3.5) constrains the dimension of the models that can be
numerically analyzed (\citet{Dushoff}). We circumvented these issues by
considering a continuous spectrum of virulence values and postulating a
model that would describe the time evolution of a continuous distribution of
virulences under the same type of competition-colonization trade-off. This
model of continuous virulence is naturally derived as the continuum limit of
the multiple-viral-strains model, thus providing a better modeling framework
for the very high phenotypic diversity of viral populations. While the model
exhibits a complicated mathematical structure as an integro-differential
Cauchy problem (\citet{Cushing}), we were able to provide a simple proof of
the existence of solutions. Having clarified the issue of existence of
solutions, we solved this Cauchy problem numerically using typical parameter
values. The discretization step size in the numerical scheme (see Appendix %
\ref{app:numerics}), clearly limits the accuracy of the numerical
approximations. However, this numerical limitation does not represent a loss
of modeling power, whereas, as explained above, computational issues do
limit the number of scenarios that can be modeled with the discrete
multiple-strains model.

Our simulation results indicate that the intra-host evolution of virulence
is characterized by two phases. During the first phase, colonizers become
more frequent and the average virulence of the population increases. In the
second phase, the abundance of competitors increases and the mean population
virulence decreases. Eventually, the virulence distribution appears to reach
a steady state in which competitors dominate strongly over colonizers.

In the two-viral-strains model (\ref{Eq.2Virs.Model}), two major assumptions
are the competition-colonization trade-off, $%
c=a_{1}^{-1}/(a_{1}^{-1}+a_{2}^{-1})$, and the cell killing rate imposed by
competitors in coinfected cells, i.e. the factor $\min (a_{1},a_{2})$ in the
last three equations of model (\ref{Eq.2Virs.Model}). The second assumption
turns out to be not crucial, which can be seen as follows. If we replace the
minimum in model (\ref{Eq.2Virs.Model}) by the maximum and assume $a_{1}\leq
a_{2}$ as before, then we obtain 
\begin{eqnarray*}
\dot{y}_{12} &=&\beta (y_{1}v_{2}+y_{2}v_{1})-a_{2}y_{12} \\
\dot{v}_{1} &=&Ka_{1}y_{1}+cKa_{2}y_{12}-uv_{1} \\
\dot{v}_{2} &=&Ka_{2}y_{2}+(1-c)Ka_{2}y_{12}-uv_{2}
\end{eqnarray*}%
for the last three equations, while all others remain unchanged. Comparing
this model to the original one we realize that strain~1 and strain~2 have
interchanged their roles in the sense that the equation for strain~1 now has
mixed virulence terms whereas the one for strain~2 has become homogeneous.
In other words, we might as well write the maximum model as%
\begin{eqnarray*}
\dot{x} &=&\lambda -dx-\beta x(v_{1}+v_{2}) \\
\dot{y}_{2} &=&\beta xv_{2}-\beta y_{2}v_{1}-a_{2}y_{2} \\
\dot{y}_{1} &=&\beta xv_{1}-\beta y_{1}v_{2}-a_{1}y_{1} \\
\dot{y}_{12} &=&\beta y_{1}v_{2}+\beta y_{2}v_{1}-a_{2}y_{12} \\
\dot{v}_{2} &=&Ka_{2}y_{2}+\widetilde{c}Ka_{2}y_{12}-uv_{2} \\
\dot{v}_{1} &=&Ka_{1}y_{1}+(1-\widetilde{c})Ka_{2}y_{12}-uv_{1}
\end{eqnarray*}%
where $\widetilde{c}:=a_{2}^{-1}/(a_{1}^{-1}+a_{2}^{-1})$. This model is
equivalent to (\ref{Eq.2Virs.Model}) with $a_{1}$ and $a_{2}$ having
switched their roles. The coexistence equilibrium in the maximum model is 
\begin{equation*}
\begin{pmatrix}
x^{\ast } \\ 
y_{2}^{\ast } \\ 
y_{1}^{\ast } \\ 
y_{12}^{\ast } \\ 
v_{2}^{\ast } \\ 
v_{1}^{\ast }%
\end{pmatrix}%
=\frac{1}{\beta K}%
\begin{pmatrix}
u \\ 
a_{1}uM/(a_{2}(M+u(a_{2}+a_{1}))) \\ 
a_{2}uM/(a_{1}(M+u(a_{2}+a_{1}))) \\ 
M^{2}/(a_{2}(M+u(a_{2}+a_{1}))) \\ 
a_{1}MK/(u(a_{2}+a_{1})) \\ 
a_{2}MK/(u(a_{2}+a_{1}))%
\end{pmatrix}%
\end{equation*}%
at which competitors are more abundant than colonizers both as free virus, $%
v_{1}^{\ast }/v_{2}^{\ast }=a_{2}/a_{1}>1$, and inside cells, $y_{1}^{\ast
}/y_{2}^{\ast }=(a_{2}/a_{1})^{2}>1$. The comparison of the minimum and the
maximum models lets us conclude that the crucial property of both models
that allows for a coexistence equilibrium in favor of competitors is not the
cell killing rate imposed, but rather the inverse proportionality between
virulence and intracellular fitness during coinfection.

To investigate the consistency of our modeling approach, we compared the
qualitative properties of the dynamics displayed by the continuous-virulence
model (\ref{Eq.Cont.Model}) and the qualitative properties of the dynamics
displayed be the discrete multiple-viral-strains model (\ref{Eq.nVirs.Model}%
) (simulation results presented in Ojosnegros et al., in preparation).
Assuming that viral spread is possible (i.e., $M>0$), we found that the
basic biphasic feature, namely, an initial phase in which the more virulent
strains colonize and expand, followed by a second phase in which the less
virulent strains predominate, is present in the dynamics of both models.
Also the shapes of the continuous distributions in the course of the
simulations show similarity to the ones observed in the discrete case, as
long as the discrete virulence values considered in the discrete model are
uniformly distributed over the interval $[a_{1},a_{2}]\subset (0,1)$ of
possible virulences. These facts are not surprising, given that the system
of ODEs solved to numerically approximate the solution of the continuous
virulence system is structurally very similar to the equations of the
discrete model, the only difference being the weights that appear in the
Newton-Cotes formulas used to approximate the integrals $%
\int_{a_{1}}^{a_{2}}v(\xi ,t)d\xi $ and $\int_{a_{1}}^{a_{2}}\frac{1}{%
a^{-1}+b^{-1}}\min (a,b)(z(a,b,t)+z(b,a,t))db$ (see Appendix~\ref%
{app:numerics}). Nevertheless, we found it very interesting that if we
simulate the continuous-virulence model using parameters such that the
condition for viral spread is not fulfilled (i.e., $M<0$), the system
evolves towards the zero density function (results not explicitly shown).
This outcome seems to be independent of the initial distributions of
virulence used. This result suggests that the condition for viral spread,
which we originally derived for the two-viral-strains model, appears to be
still correct in the continuum limit.

On the other hand, the dynamics of the two models do show an important
difference: Under viral spread conditions (i.e., $M>0$), the stationary
distributions reached in the continuous and the discrete case differ in that
the stationary continuous distribution is extremely more positively skewed
and only the very least virulent strains are represented. This finding has
unexpected implications for the discrete model (\ref{Eq.nVirs.Model}), which
will be discussed elsewhere (Ojosnegros et al., in preparation).

In conclusion, the two models studied in this article make two major
predictions about the evolution of virulence under a
competition-colonization trade-off. First, two viral strains with distinct
virulence can coexist, and second, a viral population displaying a range of
virulence values will be attenuated and evolve towards a population of many
competitors and very few colonizers. Our model predictions differ from most
prognoses based on previous models of the evolution of virulence, which
often conclude that selection maximizes the basic reproductive number of the
pathogen. This discrepancy is due to the following:

\begin{enumerate}
\item One key feature of our approach is that coinfections are explicitly
modeled.

\item We do not assume that the viral strain with the highest individual
cell killing performance dominates the events during coinfections.

\item We assume a competition-colonization trade-off.
\end{enumerate}

These three assumptions combined have not been considered in previous
modeling approaches. Under these premises, selection appears to favor
low-virulence competitors, as long as uninfected cells are constantly
replenished, but not unlimited. The attenuation property of the
continuous-virulence model may also explain experimental observations of
suppression of high-fitness viral mutants (colonizers), which might have
been displaced by competitors (%
\citet{Torre1990,Novella2004,Turner1999b,Bull2006}).

In \citet{Ojosnegros2010} two foot-and-mouth disease viral strains were
reported that had been isolated from a population undergoing viral passaging
experiments. Measurement of cell killing rates, intracellular fitness, and
other parameters suggested a competition-colonization trade-off. These
experimental findings motivate the hypothesis that viruses can specialize
either to improve their colonization skills by fast cell killing, or to
improve their competitive intracellular reproductive success. In our
modeling approach, we have implemented the competition-colonization
trade-off using the algebraically simple relationship $%
c=a_{1}^{-1}/(a_{1}^{-1}+a_{2}^{-1})$ which renders the mathematical
analysis convenient. The actual dependency between virulence $a_{i}$ and
intracellular fitness $c$ is likely to be more complicated and to depend on
additional parameters. It would be of biological interest to identify and to
characterize virus populations with a competition-colonization trade-off and
to establish the nature of the trade-off experimentally. In this article,
our principal aim was to provide a rigorous mathematical analysis of a viral
competition model incorporating a simple but archetypical instance of a
competition-colonization trade-off.

\newpage

\section*{\protect\bigskip Appendix}

\begin{appendix}
\section{Existence of solutions of the continuous-virulence model's
equations}
\label{app:existence}

In order to show that a solution of (\ref{Eq.Cont.Model}) exists, we
consider solutions during a very short time span $[t_{1},t_{2}]\subset
\lbrack 0,\infty )$ within which the values of $V(t)$ and $z(a,b,t)$ do not
significantly change, i.e. $V(t)\approx V_{t_{1}}:=V(t_{1})$ and $%
z(a,b,t)\approx z_{t_{1}}(a,b):=z(a,b,t_{1})$ $\forall $ $t\in \lbrack
t_{1},t_{2}].$ Given that (\ref{Eq.Cont.Model}) is autonomous, we may as
well consider the time interval $[t_{1}=0,t_{2}].$ Thus, $W(\theta )\approx
\int\nolimits_{0}^{\theta }d+\beta V_{0}d\tau =\theta (d+\beta V_{0})$, $%
U_{a}(\theta )\approx \int\nolimits_{0}^{\theta }a+\beta V_{0}d\tau =\theta
(a+\beta V_{0})$ $\forall $ $\theta \in \lbrack 0,t_{2}]$ and $%
\int\limits_{a_{1}}^{a_{2}}\frac{1}{a^{-1}+b^{-1}}\min
(a,b)(z(a,b,t)+z(b,a,t))db$ $\approx \int\limits_{a_{1}}^{a_{2}}\frac{1}{%
a^{-1}+b^{-1}}\min (a,b)(z_{0}(a,b)+z_{0}(b,a))db=:S_{0}(a)$ $\forall $ $%
t\in \lbrack 0,t_{2}].$ With this, equation (\ref{eq.IntegroPDEonv(a,t)})
becomes%
\begin{eqnarray*}
\frac{\partial v}{\partial t}(a,t) &=&Ka\left( y_{0}(a)+\left(
\int\limits_{0}^{t}\beta v(a,\tau )\left( x_{0}+\lambda \frac{e^{\tau
(d+\beta V_{0})}-1}{d+\beta V_{0}}\right) e^{(a-d)\tau }d\tau \right)
e^{-t(a+\beta V_{0})}\right) \\
&&+a^{-1}KS_{0}(a)-uv(a,t)
\end{eqnarray*}%
where $y_{0}(a)=y(a,0),$ $x_{0}=x(0).$ Some algebra yields%
\begin{eqnarray*}
\frac{\partial v}{\partial t}(a,t) &=&Ka\left( y_{0}(a)+\left( \beta \lambda
\int\limits_{0}^{t}\frac{\beta V_{0}}{d^{2}+d\beta V_{0}}v(a,\tau
)e^{(a-d)\tau }+\frac{1}{d+\beta V_{0}}v(a,\tau )e^{\tau (a+\beta
V_{0})}d\tau \right) e^{-t(a+\beta V_{0})}\right) \\
&&+a^{-1}KS_{0}(a)-uv(a,t) \\
&=&-uv(a,t)+K\beta \lambda ae^{-t(a+\beta V_{0})}\int\limits_{0}^{t}\left( 
\frac{\beta V_{0}}{d^{2}+d\beta V_{0}}e^{(a-d)\tau }+\frac{1}{d+\beta V_{0}}%
e^{\tau (a+\beta V_{0})}\right) v(a,\tau )d\tau \\
&&+Kay_{0}(a)+a^{-1}KS_{0}(a)
\end{eqnarray*}%
If we write $e^{-t(a+\beta V_{0})}$ as $1+(-a-\beta V_{0})t+O(t^{2})$ and
neglect terms of quadratic order we obtain for $t$ sufficiently small%
\begin{eqnarray*}
\frac{\partial v}{\partial t}(a,t) &=&-uv(a,t)+K\beta \lambda a(1+(-a-\beta
V_{0})t)\int\limits_{0}^{t}\left( \frac{\beta V_{0}}{d^{2}+d\beta V_{0}}%
e^{(a-d)\tau }+\frac{1}{d+\beta V_{0}}e^{\tau (a+\beta V_{0})}\right)
v(a,\tau )d\tau \\
&&+Kay_{0}(a)+a^{-1}KS_{0}(a)
\end{eqnarray*}%
Let us assume for a moment that a three times differentiable solution
exists. Differentiation on both sides yields%
\begin{eqnarray*}
\frac{\partial ^{2}v}{\partial t^{2}}(a,t) &=&-u\frac{\partial v}{\partial t}%
(a,t)-K\beta \lambda a(a+\beta V_{0})\int\limits_{0}^{t}\left( \frac{\beta
V_{0}}{d^{2}+d\beta V_{0}}e^{(a-d)\tau }+\frac{1}{d+\beta V_{0}}e^{\tau
(a+\beta V_{0})}\right) v(a,\tau )d\tau \\
&&+K\beta \lambda a(1+(-a-\beta V_{0})t)\left( \frac{\beta V_{0}}{%
d^{2}+d\beta V_{0}}e^{(a-d)t}+\frac{1}{d+\beta V_{0}}e^{t(a+\beta
V_{0})}\right) v(a,t)
\end{eqnarray*}%
Again, differentiation on both sides gives%
\begin{eqnarray*}
\frac{\partial ^{3}v}{\partial t^{3}}(a,t) &=&-u\frac{\partial ^{2}v}{%
\partial t^{2}}(a,t)-K\beta \lambda a(a+\beta V_{0})\left( \frac{\beta V_{0}%
}{d^{2}+d\beta V_{0}}e^{(a-d)t}+\frac{1}{d+\beta V_{0}}e^{t(a+\beta
V_{0})}\right) v(a,t) \\
&&+K\beta \lambda a(1+(-a-\beta V_{0})t)\left( \frac{\beta V_{0}}{%
d^{2}+d\beta V_{0}}e^{(a-d)t}+\frac{1}{d+\beta V_{0}}e^{t(a+\beta
V_{0})}\right) \frac{\partial v}{\partial t}(a,t) \\
&&-K\beta \lambda a(a+\beta V_{0})\left( \frac{\beta V_{0}}{d^{2}+d\beta
V_{0}}e^{(a-d)t}+\frac{1}{d+\beta V_{0}}e^{t(a+\beta V_{0})}\right) v(a,t) \\
&&+K\beta \lambda a(1+(-a-\beta V_{0})t)\left( \frac{(a-d)\beta V_{0}}{%
d^{2}+d\beta V_{0}}e^{(a-d)t}+\frac{a+\beta V_{0}}{d+\beta V_{0}}%
e^{t(a+\beta V_{0})}\right) v(a,t)
\end{eqnarray*}%
Summarizing we have%
\begin{eqnarray}
&&\frac{\partial ^{3}v}{\partial t^{3}}(a,t)+u\frac{\partial ^{2}v}{\partial
t^{2}}(a,t)-K\beta \lambda a(1+(-a-\beta V_{0})t)\left( \frac{\beta V_{0}}{%
d^{2}+d\beta V_{0}}e^{(a-d)t}+\frac{1}{d+\beta V_{0}}e^{t(a+\beta
V_{0})}\right) \frac{\partial v}{\partial t}(a,t)  \notag \\
&=&-2K\beta \lambda a(a+\beta V_{0})\left( \frac{\beta V_{0}}{d^{2}+d\beta
V_{0}}e^{(a-d)t}+\frac{1}{d+\beta V_{0}}e^{t(a+\beta V_{0})}\right) v(a,t) 
\notag \\
&&+K\beta \lambda a(1+(-a-\beta V_{0})t)\left( \frac{(a-d)\beta V_{0}}{%
d^{2}+d\beta V_{0}}e^{(a-d)t}+\frac{a+\beta V_{0}}{d+\beta V_{0}}%
e^{t(a+\beta V_{0})}\right) v(a,t)  \label{eq.ThirdOrderODE}
\end{eqnarray}

For each $a\in \lbrack a_{1},a_{2}]$ the latter equation is a third order
ordinary differential equation with variable coefficients. Using standard
Lipschitz-continuity arguments (see, for instance, Section 4.3 in %
\citet{Koenigsberger}) it can be shown that for each $a\in \lbrack
a_{1},a_{2}]$ the initial value problem (\ref{eq.ThirdOrderODE}) together
with $v(a,0)=v_{0}(a),\frac{\partial v}{\partial t}%
(a,0)=-uv_{0}(a)+Kay_{0}(a)+a^{-1}KS_{0}(a)$ and\linebreak\ $\frac{\partial
^{2}v}{\partial t^{2}}(a,0)=-u\left(
-uv_{0}(a)+Kay_{0}(a)+a^{-1}KS_{0}(a)\right) +K\beta \lambda a\left( \frac{%
\beta V_{0}}{d^{2}+d\beta V_{0}}+\frac{1}{d+\beta V_{0}}\right) v_{0}(a)$
(which we assume to be continuous functions of $a$) must have a unique
solution defined on some interval $[0,T_{1}]\subset \lbrack 0,\infty )$ of
positive length $T_{1}\in 
\mathbb{R}
_{+}.$ Given that the coefficients of (\ref{eq.ThirdOrderODE}) are
continuous functions of $a$ (which can be interpreted as a parameter in the
ODE (\ref{eq.ThirdOrderODE}) in the framework of a sensitivity analysis) all
the solutions $v(a,t)$ must be continuous on $[a_{1},a_{2}]\times \lbrack
0,T_{1}]$ (see, for instance, Theorem 6.1 in \citet{ODEnumerics} and also
Subsection 3.1.1 in \citet{Bornemann}). This family of solutions allows us
to construct a local solution (defined on $[a_{1},a_{2}]\times \lbrack
0,T_{1}]$) of (\ref{Eq.Cont.Model}) by means of substitution of $v(a,t)$
into the expressions (\ref{Eq.Exp.ForX}), (\ref{Eq.Exp.ForY}) and (\ref%
{Eq.Exp.ForZ}). The procedure can be now repeated for a short time interval
starting at $T_{1}$ yielding the next local solution. A global solution can
be obtained by patching together the local solutions and letting the $%
T_{i}\rightarrow 0$.

\section{Numerical solution of the continuous-virulence model's equations}
\label{app:numerics}

To solve the system of equations (\ref{Eq.Cont.Model}) numerically, we
discretized the "virulence-space" with an equidistant grid $%
G_{n}([0.01,0.5]) $ and approximated the integrals $\int_{a_{1}}^{a_{2}}v(%
\xi ,t)d\xi \approx \sum_{j\in G_{n}([0.01,0.5])}\gamma _{j}v(j,t)$ and $%
\int_{a_{1}}^{a_{2}}\frac{1}{a^{-1}+b^{-1}}\min
(a,b)(z(a,b,t)+z(b,a,t))db\approx \sum_{j\in G_{n}([0.01,0.5])}\gamma _{j}%
\frac{1}{a^{-1}+j^{-1}}\min (a,j)(z(a,j,t)+z(j,a,t))$ using a Newton-Cotes
quadrature formula of seventh order (with weights $\gamma _{j}\,$). After
this discretization step, we obtain for each pair $(a,b)\in
(G_{n}([0.01,0.5])\times G_{n}([0.01,0.5]))$ the following system of
ordinary differential equations%
\begin{eqnarray*}
\dot{x}(t) &=&\lambda -dx(t)-\beta x(t)\sum_{j\in G_{n}([0.01,0.5])}\gamma
_{j}v(j,t) \\
\frac{dy}{dt}(a,t) &=&\beta x(t)v(a,t)-\beta y(a,t)\sum_{j\in
G_{n}([0.01,0.5])}\gamma _{j}v(j,t)-ay(a,t) \\
\frac{dz}{dt}(a,b,t) &=&\beta y(a,t)v(b,t)-\min (a,b)z(a,b,t) \\
\frac{dv}{dt}(a,t) &=&Kay(a,t)+a^{-1}K\left( \sum_{j\in
G_{n}([0.01,0.5])}\gamma _{j}\frac{1}{a^{-1}+j^{-1}}\min
(a,j)(z(a,j,t)+z(j,a,t))\right) -uv(a,t)
\end{eqnarray*}%
The resulting system of coupled ordinary differential equations is solved
numerically.

\section{Detailed description of Figures 2 and 3}
\label{app:figures2&3}

The time evolution of an initial Gaussian distribution of virulences is
shown in snapshots in Figure 2. Initially we observe amplification, followed
by even stronger amplification in combination with slight right-shift. At $%
t=20.2$ $h$ this phase reaches a peak. A minor recession follows and
left-shifting sets on. At $t=700$ $h$ left shifting has progressed and
moderate amplification sets on again. As left-shifting continues the
distribution becomes narrower and narrower. The dynamics become slower as
the left traveling peak approaches the left boundary of the interval $%
[0.01,0.5].$ Once it reaches the boundary, the distribution starts loosing
its Gaussian shape to become an exponentially shaped distribution in favor
of the least virulent part. The changes become very slow and, at least
numerically, the system seems to be reaching a stationary distribution,
which is exponential and highly in favor of the smallest virulence values.

The time evolution of an initial Gaussian mixture distribution of virulences
is shown in snapshots in Figure 3. Initially we observe amplification, with
clear advantage for the most virulent component of the Gaussian mixture. At $%
t=17.19$ $h$ this component reaches a peak, while the least virulent
component continues growing (order of magnitude $10^{7},$ thus not visible
on the scale depicted). After this point, recession sets on. At $t=22.23$ $h$
both components are shrinking, the least virulent one only slightly (not
visible). After approximately 14 hours the least virulent component starts
growing again, while the most virulent one continues shrinking. By $t=178.15$
$h$ we can already see the least virulent component reach the order of
magnitude of $10^{7}.$ By $t=446.52$ the most virulent component is about to
fall below this order of magnitude. The least virulent component continues
growing while the most virulent one declines further. At $t=9600$ $h$ the
most virulent component has vanished (numerical order of magnitude $%
10^{-104} $) and the least virulent one starts shifting to the left while
further amplified. At $t=37600$ $h$ it becomes visible that the least
virulent component starts also narrowing, while growing and left-shifting
continues. The dynamics remain qualitatively the same, become significantly
slower though. The changes become very slow and, at least numerically, the
system seems to be reaching a stationary distribution with mode 0.0393. This
value is approximately $38\times \sigma _{1}$ left from $\mu _{1}.$

\end{appendix}

\bibliographystyle{spmpscinat}
\bibliography{biology,MathRef,nikos-dob}

\begin{thebibliography}{31}
\providecommand{\natexlab}[1]{#1}
\providecommand{\url}[1]{#1}
\providecommand{\urlprefix}{URL }
\expandafter\ifx\csname urlstyle\endcsname\relax
  \providecommand{\doi}[1]{DOI~\discretionary{}{}{}#1}\else
  \providecommand{\doi}{DOI~\discretionary{}{}{}\begingroup
  \urlstyle{rm}\Url}\fi

\bibitem[{Barnett and {\v{S}}iljak(1977)}]{OnRouthsCriterion}
Barnett, S., {\v{S}}iljak, D.D.: Routh's algorithm: a centennial survey.
\newblock SIAM Rev. \textbf{19}(3), 472--489 (1977)

\bibitem[{Bonhoeffer et~al.(1996)Bonhoeffer, Lenski, and Ebert}]{LenskiII}
Bonhoeffer, S., Lenski, R.E., Ebert, D.: The curse of the pharaoh: The
  evolution of virulence in pathogens with long living propagules.
\newblock Proceedings: Biological Sciences \textbf{263}(1371), 715--721 (1996).
\newblock \urlprefix\url{http://www.jstor.org/stable/50702}

\bibitem[{Bonhoeffer et~al.(1997)Bonhoeffer, May, Shaw, and Nowak}]{Bonhoeffer}
Bonhoeffer, S., May, R.M., Shaw, G.M., Nowak, M.A.: Virus dynamics and drug
  therapy.
\newblock Proceedings of the National Academy of Sciences of the United States
  of America \textbf{94}(13), 6971--6976 (1997).
\newblock \urlprefix\url{http://www.pnas.org/content/94/13/6971.abstract}

\bibitem[{Bull(1994)}]{BullVirulenceAReview}
Bull, J.J.: Perspective: Virulence.
\newblock Evolution \textbf{48}(5), 1423--1437 (1994).
\newblock \urlprefix\url{http://www.jstor.org/stable/2410237}

\bibitem[{Bull et~al.(2006)Bull, Millstein, Orcutt, and Wichman}]{Bull2006}
Bull, J.J., Millstein, J., Orcutt, J., Wichman, H.A.: Evolutionary feedback
  mediated through population density, illustrated with viruses in chemostats.
\newblock Am Nat \textbf{167}(2), E39--E51 (2006).
\newblock \doi{10.1086/499374}.
\newblock \urlprefix\url{http://dx.doi.org/10.1086/499374}

\bibitem[{Cushing(1977)}]{Cushing}
Cushing, J.M.: Integrodifferential equations and delay models in population
  dynamics.
\newblock Lecture notes in Biomathematics. Springer-Verlag, New York (1977)

\bibitem[{Deuflhard and Bornemann(2008)}]{Bornemann}
Deuflhard, P., Bornemann, F.: Numerische Mathematik 2.
\newblock de Gruyter Lehrbuch. [de Gruyter Textbook], revised edn. Walter de
  Gruyter \& Co., Berlin (2008).
\newblock Gew{\"o}hnliche Differentialgleichungen. [Ordinary differential
  equations]

\bibitem[{Domingo and Holland(1997)}]{Domingo1997}
Domingo, E., Holland, J.J.: {RNA} virus mutations and fitness for survival.
\newblock Annu Rev Microbiol \textbf{51}, 151--178 (1997).
\newblock \doi{10.1146/annurev.micro.51.1.151}.
\newblock \urlprefix\url{http://dx.doi.org/10.1146/annurev.micro.51.1.151}

\bibitem[{Eigen et~al.(1988)Eigen, McCaskill, and Schuster}]{Eigen1988}
Eigen, M., McCaskill, J., Schuster, P.: Molecular quasi-species.
\newblock J Phys Chem \textbf{92}(24), 6881--6891 (1988)

\bibitem[{Epperson(2007)}]{ODEnumerics}
Epperson, J.F.: An introduction to numerical methods and analysis.
\newblock Revised edn. Wiley-Interscience [John Wiley \& Sons], Hoboken, NJ
  (2007)

\bibitem[{Haraguchi and Sasaki(2000)}]{EvoVirII}
Haraguchi, Y., Sasaki, A.: The evolution of parasite virulence and transmission
  rate in a spatially structured population.
\newblock Journal of Theoretical Biology \textbf{203}(2), 85 -- 96 (2000).
\newblock \doi{DOI: 10.1006/jtbi.1999.1065}.
\newblock
  \urlprefix\url{http://www.sciencedirect.com/science/article/B6WMD-45KV7C6-T/%
2/51ed4b299f5aab5ac70052e24153ee50}

\bibitem[{Herrera et~al.(2007)Herrera, Garcia-Arriaza, Pariente, Escarmis, and
  Domingo}]{EstebanDomingo}
Herrera, M., Garcia-Arriaza, J., Pariente, N., Escarmis, C., Domingo, E.:
  Molecular basis for a lack of correlation between viral fitness and cell
  killing capacity.
\newblock PLoS Pathog \textbf{3}(4), e53 (2007).
\newblock \doi{10.1371/journal.ppat.0030053}

\bibitem[{Holland et~al.(1992)Holland, Torre, and Steinhauer}]{Holland1992}
Holland, J.J., Torre, J.C.D.L., Steinhauer, D.A.: R{NA} virus populations as
  quasispecies.
\newblock Curr Top Microbiol Immunol \textbf{176}, 1--20 (1992)

\bibitem[{{Jung} et~al.(2002){Jung}, {Maier}, {Vartanian}, {Bocharov}, {Jung},
  {Fischer}, {Meese}, {Wain-Hobson}, and {Meyerhans}}]{CoinfectionSize}
{Jung}, A., {Maier}, R., {Vartanian}, J., {Bocharov}, G., {Jung}, V.,
  {Fischer}, U., {Meese}, E., {Wain-Hobson}, S., {Meyerhans}, A.:
  Recombination: Multiply infected spleen cells in hiv patients.
\newblock Nature \textbf{418}, 144--+ (2002)

\bibitem[{K{\"o}nigsberger(2004)}]{Koenigsberger}
K{\"o}nigsberger, K.: Analysis 2.
\newblock 5th edn. Springer, Berlin, Heidelberg (2004)

\bibitem[{Korobeinikov(2004)}]{Korobeinikov}
Korobeinikov, A.: Global properties of basic virus dynamics models.
\newblock Bull. Math. Biol. \textbf{66}(4), 879--883 (2004)

\bibitem[{Kryazhimskiy et~al.(2007)Kryazhimskiy, Dieckmann, Levin, and
  Dushoff}]{Dushoff}
Kryazhimskiy, S., Dieckmann, U., Levin, S.A., Dushoff, J.: On state-space
  reduction in multi-strain pathogen models, with an application to antigenic
  drift in influenza a.
\newblock PLoS Comput Biol \textbf{3}(8), e159 (2007).
\newblock \doi{10.1371/journal.pcbi.0030159}

\bibitem[{Lenski(1988)}]{LenskiIV}
Lenski, R.E.: Evolution of plague virulence.
\newblock Nature \textbf{334}, 473--474 (1988)

\bibitem[{Lenski and Levin(1985)}]{LenskiV}
Lenski, R.E., Levin, B.R.: Constraints on the coevolution of bacteria and
  virulent phage: A model, some experiments, and predictions for natural
  communities.
\newblock The American Naturalist \textbf{125}(4), 585--602 (1985).
\newblock \urlprefix\url{http://www.jstor.org/stable/2461275}

\bibitem[{Lenski and May(1994)}]{LenskiIII}
Lenski, R.E., May, R.M.: The evolution of virulence in parasites and pathogens:
  Reconciliation between two competing hypotheses.
\newblock Journal of Theoretical Biology \textbf{169}(3), 253 -- 265 (1994).
\newblock \doi{DOI: 10.1006/jtbi.1994.1146}.
\newblock
  \urlprefix\url{http://www.sciencedirect.com/science/article/B6WMD-45NJH8K-3J%
/2/9d01adae902dd4f3542ef062c5055e64}

\bibitem[{May and Nowak(1995)}]{CoinfectionAndEvolutionVir}
May, R.M., Nowak, M.A.: Coinfection and the evolution of parasite virulence.
\newblock Proceedings: Biological Sciences \textbf{261}(1361), 209--215 (1995).
\newblock \urlprefix\url{http://www.jstor.org/stable/50287}

\bibitem[{Novella et~al.(2004)Novella, Reissig, and Wilke}]{Novella2004}
Novella, I.S., Reissig, D.D., Wilke, C.O.: Density-dependent selection in
  vesicular stomatitis virus.
\newblock J Virol \textbf{78}(11), 5799--5804 (2004).
\newblock \doi{10.1128/JVI.78.11.5799-5804.2004}.
\newblock \urlprefix\url{http://dx.doi.org/10.1128/JVI.78.11.5799-5804.2004}

\bibitem[{Nowak and May(2000{\natexlab{a}})}]{Nowak2000}
Nowak, M., May, R.: Virus dynamics.
\newblock Oxford University Press (2000{\natexlab{a}})

\bibitem[{Nowak and May(2000{\natexlab{b}})}]{NowakMay}
Nowak, M.A., May, R.M.: Virus dynamics.
\newblock Oxford University Press, Oxford (2000{\natexlab{b}}).
\newblock Mathematical principles of immunology and virology

\bibitem[{Ojosnegros et~al.(2010)Ojosnegros, Beerenwinkel, Antal, Nowak,
  Escarmísa, and Domingo}]{Ojosnegros2010}
Ojosnegros, S., Beerenwinkel, N., Antal, T., Nowak, M.A., Escarmísa, C.,
  Domingo, E.: Competition-colonization dynamics in an {RNA} virus.
\newblock Proc Natl Acad Sci U S A p. in press (2010)

\bibitem[{O'Keefe and Antonovics(2002)}]{EvoVirI}
O'Keefe, K.J., Antonovics, J.: Playing by different rules: The evolution of
  virulence in sterilizing pathogens.
\newblock The American Naturalist \textbf{159}(6), 597--605 (2002).
\newblock \urlprefix\url{http://www.jstor.org/stable/3079040}

\bibitem[{Perelson and Nelson(1999)}]{Perelson1999}
Perelson, A., Nelson, P.: Mathematical analysis of {HIV-1} dynamics in vivo.
\newblock SIAM Review \textbf{41}(1), 3--44 (1999)

\bibitem[{Taylor et~al.(1998)Taylor, Jarosz, Lenski, and Fulbright}]{LenskiI}
Taylor, D.R., Jarosz, A.M., Lenski, R.E., Fulbright, D.W.: The acquisition of
  hypovirulence in host-pathogen systems with three trophic levels.
\newblock The American Naturalist \textbf{151}(4), 343--355 (1998).
\newblock \urlprefix\url{http://www.jstor.org/stable/2463421}

\bibitem[{Tilman(2007)}]{Tilman2007}
Tilman, D.: Theoretical Ecology: Principles and Applications, chap.
  Interspecific competition and multispecies coexistence, pp. 84--97.
\newblock Oxford University Press (2007)

\bibitem[{de~la Torre and Holland(1990)}]{Torre1990}
de~la Torre, J.C., Holland, J.J.: {RNA} virus quasispecies populations can
  suppress vastly superior mutant progeny.
\newblock J Virol \textbf{64}(12), 6278--6281 (1990)

\bibitem[{Turner and Chao(1999)}]{Turner1999b}
Turner, P.E., Chao, L.: Prisoner's dilemma in an {RNA} virus.
\newblock Nature \textbf{398}(6726), 441--443 (1999).
\newblock \doi{10.1038/18913}.
\newblock \urlprefix\url{http://dx.doi.org/10.1038/18913}

\end{thebibliography}

\end{document}